\documentclass[aps,prl,twocolumn,groupedaddress]{revtex4}

\usepackage{graphicx}% Include figure files
\usepackage{dcolumn}% Align table columns on decimal point
\usepackage{bm}% bold math
\usepackage{amssymb}
\usepackage{amsmath}
\usepackage{textcomp}

\usepackage{color}

\begin{document}

\title{Emergence and Control of Complex Behaviours in Driven Systems of Interacting Qubits with Dissipation}

\author{A. V.  Andreev$^1$,  A. G. Balanov$^{2,+}$, T. M. Fromhold$^3$, M. T. Greenaway$^2$, A. E. Hramov$^1$, W. Li$^3$,  V.  V. Makarov$^1$, A. M. Zagoskin$^{2,4,*}$ }
\affiliation{$^1$Innopolis University, Universitetskaya Str. 1, Innopolis, 420500, Russia}
\affiliation{$^2$Department of Physics, Loughborough University, Loughborough LE11 3TU, United Kingdom}
\affiliation{$^3$School of Physics and Astronomy, University of Nottingham, Nottingham NG7 2RD, United Kingdom}
\affiliation{$^4$Department of Theoretical Physics and Quantum Technologies, National University of Science and Technology MISIS, 4 Leninsky Ave., Moscow 119049, Russia}
\affiliation{$^*$Corresponding author: a.zagoskin@lboro.ac.uk}
\affiliation{$^+$Corresponding author: a.balanov@lboro.ac.uk}
\date{\today}

\begin{abstract} Progress in the creation of large scale, artificial quantum coherent structures demands the investigation of their nonequilibrium dynamics when strong  interactions, even between remote parts, are non-perturbative. Analysis of multiparticle quantum correlations in a large system in the presence of decoherence and external driving is especially topical. Still, the scaling behaviour of dynamics and related emergent phenomena are not yet well understood.  We investigate how the dynamics of a driven system of several quantum elements (e.g., qubits or Rydberg atoms) changes with increasing number of elements.  Surprisingly, a two-element  system exhibits chaotic behaviours. For larger system sizes a highly stochastic, far from equilibrium, {\em hyperchaotic} regime emerges. Its complexity systematically scales with the size of the system,   proportionally to the number of elements. Finally, we demonstrate that these chaotic dynamics can be efficiently controlled by a periodic driving field. The insights provided by our results indicate the possibility of  a reduced description  for the behaviour of a large quantum system in terms of the transitions between its qualitatively different dynamical regimes. These transitions are controlled by a relatively small number of parameters, which may prove useful in the design, characterization and control of large artificial quantum structures.
\end{abstract}

 \maketitle

\section{Introduction}

Recent progress in experimental techniques for the fabrication, control and measurement of quantum coherent systems has allowed the routine creation of moderate-to-large arrays of controllable quantum units (e.g., superconducting qubits, trapped atoms and ions) expected to have revolutionary applications, e.g., in sensing, quantum communication and quantum computing  \cite{Walport2016,Georgescu2012,Knight2019,Acin2018}.  They also allow the investigation of fundamental physics, such as the measurement problem \cite{Zurek2003,Zurek2009,Xiong2015}, the link between the chaotic behaviour in classical systems and the corresponding dynamics in their quantum counterparts \cite{Lambert2009,assmann2016quantum,fiderer2018quantum,Gao2015,Zhu2019} and the relative roles of entanglement and thermalization \cite{Chaud2009,Zhang2010,Neil2016,Turner2018,Lewis2019}. 

The deployment of modern quantum technologies generally requires both the improvement of unit quantum elements (e.g., increasing their decoherence times) and the scaling up of quantum structures \cite{Dowling2003}. 

It is known that large systems comprising many functional elements tend to demonstrate emergent phenomena in their dynamics, which can not be observed in smaller systems \cite{Hegel2015,Engels1969}.  Examples include dynamical phases in driven Bose-Einstein condensates \cite{Piazza15}, synchronization in micromechanical oscillator arrays \cite{Zhang15} and coherent THz emission from layered high-temperature superconductors \cite{Ozyuzer07,Saveliev2010,Welp13}. Nevertheless, the emergent phenomena are commonly neglected when designing and investigating large-scale coherent structures. This is possibly due to the absence of a convenient theoretical framework for studying these novel systems.  One important open question to be tackled is how the dynamics scales with the increase of the system size (e.g., number of qubits), and what role in its change is played by emergent phenomena.  

We address these questions by studying theoretically the far-from-equilibrium dynamics of a driven chain of interacting two-level quantum systems with dissipation and dephasing [Fig. 1a]. We find that even a short chain comprising between two and four elements can demonstrate chaotic behaviour. Remarkably, in systems with five or more elements a phenomenon known as hyperchaos emerges, a dynamical regime characterized by two or more positive Lyapunov exponents. Hyperchaos has features that resemble thermalisation even though the system is far from equilibrium. We show that, in contrast to thermalisation,  the deterministic origin of hyperchaos allows it to be controlled either by tuning of the system parameters, or by applying an external driving force.  Our results give new insights into the dynamics of large artificial quantum coherent structures,  which will be important for the design and control of quantum systems such as Rydberg molecules, quantum information processors, simulators and detectors.

\section{Results}
\subsection{Mathematical model}

\begin{figure*}[t]
	%\vskip 4cm
	\centerline{\includegraphics*[scale=.3]{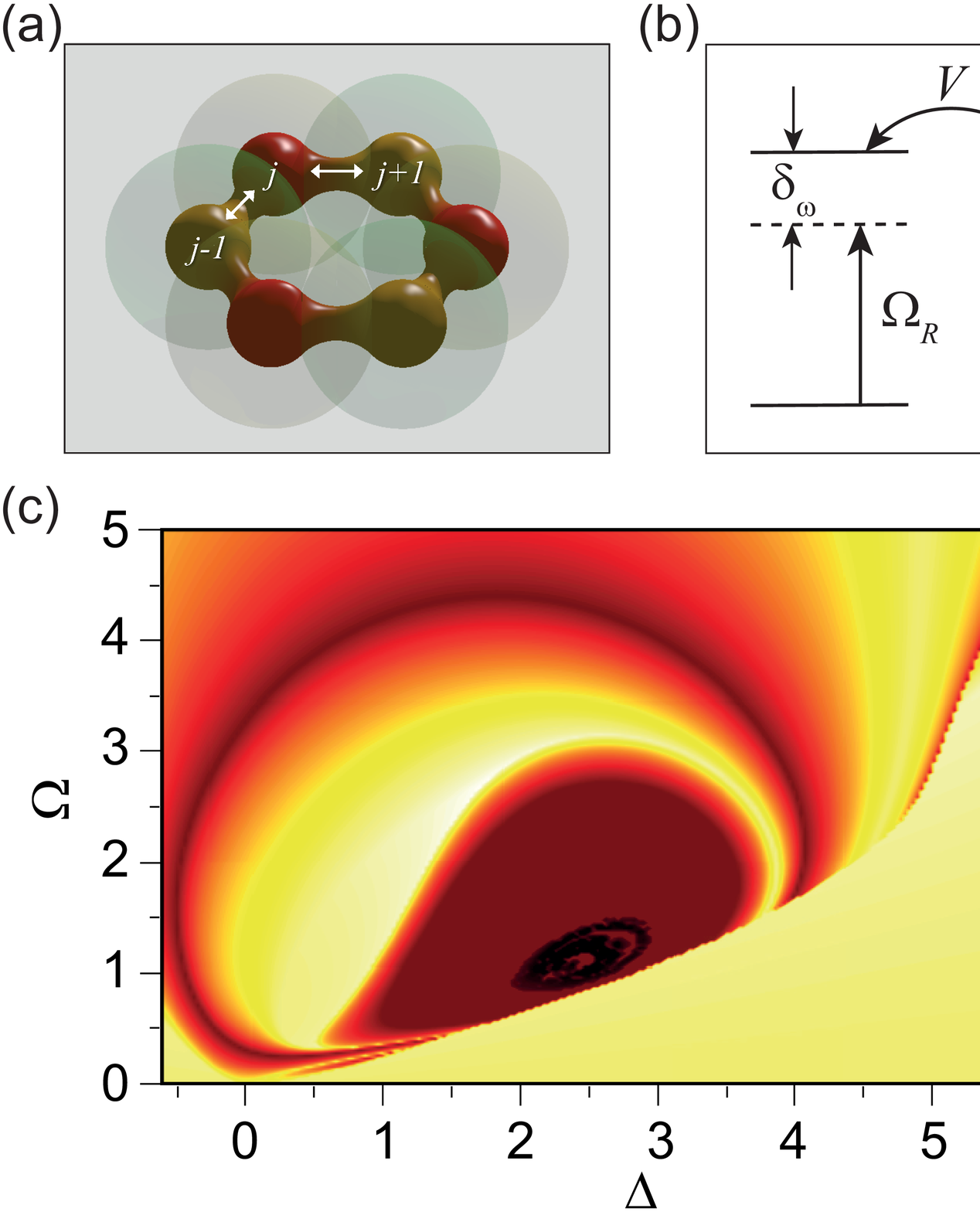}}
	\caption{\textbf{
			%Model of interacting quantum units + Lyapunov exponent diagram for 2 qubits.
			Schematic representation of interacting qubits and the largest Lyapunov exponent dependence on model parameters for the case of two qubits.} \textbf{a} Schematic representation of a circular chain of six interacting quantum units (e.g., Rydberg atoms or qubits), labelled by integer indexes and coupled via nearest neighbor interaction. \textbf{b} Atomic-level configuration reflected in the model [Eqs. (\ref{eq:Ham}-\ref{eq:Lin})]. Coherent laser excitation from the ground state $|g\rangle$ to an excited ("Rydberg") state $|e\rangle$ with a Rabi frequency $\Omega_R$ and detuning $\delta_{\omega}$, incoherent spontaneous decay $\gamma$, and the Rydberg interaction when an excited qubit shifts the transition frequency of a neighboring qubit by $V$. \textbf{c} Dependence of the largest Lyapunov exponent $\Lambda_1$ on $\Delta$ and $\Omega$ calculated for $c=5$ for the system of 2 coupled qubits. The color scale intensity indicates the value of $\Lambda_1$.}
	\label{fgr:Model}
\end{figure*}

Here, we investigate  the dynamics of a 1D chain of $N$  qubits driven by an external electromagnetic field 
%(with Rabi frequency $\Omega_r$ and detuning $\delta_w$)
in the presence of decoherence. The system under consideration is equivalent to a system of Rydberg atoms, by rescaling and reidentification of parameters ~\cite{Weimer:2010aa,Lee2011,Lee2012,Ostrovskaya2016,Labuhn:2016aa,Minganti2019}. Along with superconducting structures,  Rydberg systems provide a controllable testbed for studying fundamental quantum dynamics~\cite{Bendkowsky:2009aa,Low:2014aa,basak_periodically_2018} and for developing new quantum chemistry \cite{Butscher:2010aa,Niederprum:2016aa,Eiles:2018aa}. In particular, arrays of Rydberg atoms can be controlled experimentally with very  high precision~\cite{zeiher_coherent_2017,labuhn_tunable_2016,barredo_coherent_2015}. Each atom has a ground state $|g\rangle$ and an electronically excited high-energy Rydberg state $|e\rangle$ [Fig.~\ref{fgr:Model}b]. Rydberg states have large polarizability $\sim n^{7}$ (where $n$ is the principal quantum number), which generates strong and long-range interactions between atoms~\cite{RevModPhys.82.2313}. 

To model an electromagnetically driven array of qubits (Rydberg atoms), we use the Hamiltonian
\begin{eqnarray}
H = \sum_{j=1}^N \left[ -\delta_{\omega} |e\rangle\langle e|_j + \frac{\Omega_R}{2}\left(|e\rangle\langle g|_j+|g\rangle\langle e|_j\right)\right] 
\label{eq:Ham}
\\ + \sum_{j<k}V_{i,j} |e\rangle \langle e|_j \otimes |e\rangle\langle e|_k \nonumber, 
\end{eqnarray}
where $\delta_{\omega} = \omega_l-\omega_0$ is the detuning between the laser and transition frequencies, $\Omega_R$ is the Rabi frequency (tuned by the laser field amplitude), and $V_{i,j}$ characterizes the interaction between $i$th and $j$th qubits. 

The dynamics are described by the Liouville-von Neumann equation for the density matrix $\rho$, 
\begin{equation}
\dot{\rho}=-i\left[H,\rho\right]+\mathcal{L}\left[\rho\right],
\label{eq:rho}
\end{equation}
where relaxation and dephasing processes are taken into account through the appropriate Lindblad operator
\begin{equation}
\mathcal{L}[\rho]=\gamma\sum_j\left[ |g\rangle\langle e|_j \rho |e\rangle\langle g|_j-\frac{1}{2}\{|e\rangle\langle e|_j,\rho\} \right],
\label{eq:Lin}
\end{equation}
with $\gamma$ to be the decay rate from the excited state to the ground state. Rydberg atom interactions can range over a few tens of micrometers, larger than the physical size of the system. (The same is true for superconducting qubits interacting through a resonator field mode).  Therefore, to simplify the analysis, we can initially assume that interactions are identical for any pair of qubits, $V_{i,j}= V$  [see Fig.~\ref{fgr:Model}a,b]. For a system of $N$  qubits, $\rho$ has the dimensionality of its Hilbert space, $2^N$. 

This problem can be greatly simplified by substituting a fully factorized density matrix approximation,
$\rho \approx \prod_j\otimes \rho_j$, in (\ref{eq:rho}). Here $\rho_j$ is the (one-particle) density matrix of the $j$th qubit.
This approximation is justified for a partially quantum coherent system, for which it has been shown accurately to describe experimental measurements on qubits, up to the corrections due to two-point correlations \cite{Zagoskin2009a,Macha2014, Gaul2016, Zeiher2016, zeiher_coherent_2017}.  In such a case, the quantum coherence between spatially separated elements is disrupted, e.g. by local ambient noise \cite{Aolita2015}.

Rewriting the factorized density matrix $\rho$ in terms of the population inversion of the $j$th qubit, $w_j = (\rho_j)_{11} - (\rho_j)_{00}$, and its coherence, $q_j = (\rho_j)_{10} = (\rho_j)_{01}^*$, yields:
\begin{eqnarray}
\dot{w}_j &=& -2\Omega\Im q_j - (w_j+1);\nonumber \\
\dot{q}_j &=& i\left[\Delta-c\sum_{k\neq j}(w_k+1) \right]q_j - \frac{1}{2}q_j + i\frac{\Omega}{2} w_j. 
\label{eq:3}
\end{eqnarray}
In (\ref{eq:3})  the dimensionless time $\tau=\gamma t$ (i.e. $\dot{x}\equiv dx/d\tau = \gamma^{-1}dx/dt$),  $\Omega=\Omega_R/\gamma$, $\Delta=\delta_{\omega}/\gamma$, and $c=Va/\gamma$, where $a$ is the lattice constant. In more detail the derivation of (\ref{eq:3}) is discussed in Section A in Supplementary Materials.

\subsection{Emergence of chaos and hyperchaos}

Here we study the dynamical regimes of chains of interacting qubits. To characterise the complexity (randomness) of the emergent chaos and hyperchaos we calculate the full spectrum of the Lyapunov exponents \cite{AnishchBook95}.

First we analyze two coupled qubits ($N=2$). Previously, it has been shown that interplay between the energy pumping and dissipation can eventually trigger self-sustained state population oscillations in this system and even lead to the emergence of bistability, when homogeneous and antiferromagnetic states coexist \cite{Lee2011}. Our investigation reveals another interesting phenomenon associated with deterministic chaos, which emerges via a cascade of period-doubling bifurcations for a periodic oscillations  \cite{AnishchBook95}.
In Fig. \ref{fgr:Model}c, we show a color map illustrating the dependence of the largest Lyapunov exponent $\Lambda_1$ on  $\Delta$ and $\Omega$. %The negative values of $\Lambda_1$ correspond to the fixed points (equilibrium states), zero values mean the periodic or quasiperiodic solutions, while positive $\Lambda_1$ evidences the presence of deterministic chaos \cite{AnishchBook95}. 
A  transition from  yellow to black reflects the change from smaller to larger values of $\Lambda_1$. The diagram clearly demonstrates the presence of chaotic dynamics, for which $\Lambda_1>0$, in considerable areas of the parameter plane shaded by  black color.
This shows that the discovered chaos is a robust phenomenon existing in our system over wide parameter ranges. Notably, for certain parameter values it coexists with the antiferromagnetic steady state. The detailed descriptions of the  bifurcation transitions and analysis of the Lyapunov exponents that characterize the stability of long-term dynamics are presented in Section B in Supplementary Materials.

%Supplemental materials \cite{Bal19supp}. 

\begin{figure*}[t]
	\centerline{\includegraphics*[scale=.4]{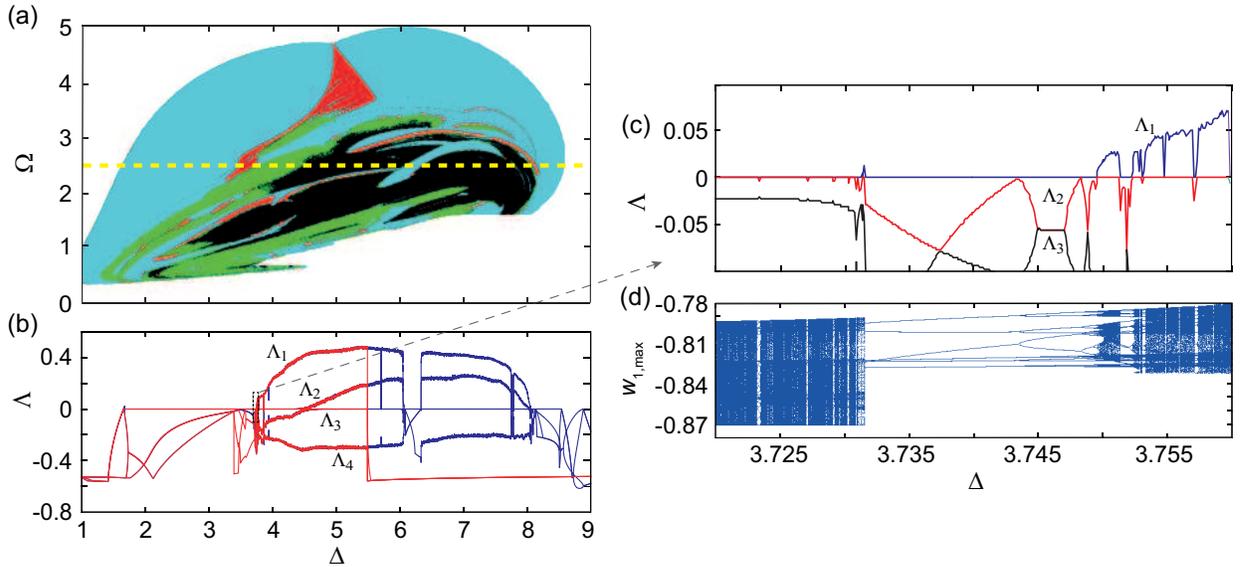}}
	\caption{\textbf{Hyperchaos in a ring of five  qubits.} \textbf{a} Map showing areas of ($\Delta$, $\Omega$) space corresponding to different regimes of long-term behavior for a ring of five qubits. Black denotes areas of hyperchaos with two positive Lyapunov exponents, green is a region of chaos with one positive Lyapunov exponent, cyan is a periodic regime, red areas correspond to quasiperiodic behavior, white is a steady state regime. \textbf{b} Variation of the four largest Lyapunov exponents with $\Delta$; blue dots correspond to $\Delta$ increasing from 1 to 9, red dots  -- to $\Delta$ decreasing down to 1. \textbf{c} A zoom of the region of \textbf{b} within the black rectangle. \textbf{d} Single-parametric bifurcation diagram corresponding to \textbf{c}, in which the vertical positions of each point correspond to the local maxima of $w_1(t)$. For all figures $\Omega=2.5$.}
	\label{fgr:map}
\end{figure*}

%In order to check whether the chaos survives in larger chains, we have examined the emergent dynamical regimens for larger numbers $N$ of qubits in the ring.
For $N=$ 3 and 4, the same type of chaotic dynamics exists in the system. 
However, surprisingly,  for $N\geq 5$ a new type of chaotic dynamics appears, whose stability is characterized by more than one positive Lyapunov exponent. This type of chaos is known as {\it hyperchaos} \cite{Roes_Hyper79}. A map of different dynamical regimes in the ($\Delta$, $\Omega$) parameter plane is shown in Fig.~\ref{fgr:map}a for $N=5$ and $c=$ 5. This diagram was built by calculating the spectrum of Lyapunov exponents for the various limit sets that exist in the ($\Delta$, $\Omega$)-plane. When all the Lyapunov exponents are negative, there is a stable equilibrium state (white in the diagram).  If the  largest Lyapunov exponent equals 0  the oscillations are periodic (cyan).  If the largest two Lyapunov exponents are both 0, the dynamics are quasiperiodic (red). However, when one or two Lyapunov exponents are positive, there is chaos (green) or hyperchaos (black), respectively.

%To illustrate the feature of hyperchaos, we show the bifurcation transitions between different regimes for $\Omega=2.5$ in Fig. \ref{fgr:1ddia} (a). This corresponds to changing $\Delta$  along the dashed line in Fig. \ref{fgr:map}. 

Figure \ref{fgr:map}b illustrates the bifurcation transitions between the different dynamical regimes  when $\Omega=2.5$ and $\Delta$ changes
along the yellow dashed line in Fig. \ref{fgr:map}a. 
The blue curves represent the evolution of the four largest Lyapunov exponents $\Lambda_1>$$\Lambda_2>$$\Lambda_3>$$\Lambda_4$. A number of distinctive phases emerge as we increase $\Delta$ from 1 to 9. The stable equilibrium state, which exists for small $\Delta$, switches to a periodic solution as a result of an Andronov-Hopf bifurcation at $\Delta \approx1.7$, where $\Lambda_1$ becomes zero. At $\Delta\approx3.55$ the periodic oscillations lose their stability via a Neimark-Sacker bifurcation, resulting in the onset of quasiperiodic oscillations ($\Lambda_1=$ $\Lambda_2=0$). Increasing  $\Delta$ further leads to chaotic dynamics.

To better illustrate the transition to the chaotic regime, in Fig. \ref{fgr:map}c we present a zoom of the region of Fig. \ref{fgr:map}b framed by the black rectangle. The corresponding bifurcation diagram, shown in Fig. \ref{fgr:map}d, is constructed by plotting the points corresponding to the local maxima  $w_{1,max}$ in the time evolution of  $w_1(t)$, calculated for given $\Delta$. For a particular value of $\Delta$, periodic solutions are represented by one or few single dots on the graph, while the complex sets of many points for a specific $\Delta$ reflect quasiperiodic or chaotic dynamics. As $\Delta$ increases, the quasiperiodic oscillations where $\Lambda_1= $$\Lambda_2=0$ [Fig. \ref{fgr:map}c] are replaced (at $\Delta\approx3.732$) by  complex periodic oscillations due to saddle-node bifurcation. These periodic oscillations are characterized by $\Lambda_1=0$  and $\Lambda_{2,3}<0$ [Fig. \ref{fgr:map}c] and represented in Fig. \ref{fgr:map}d by few isolated dots for a fixed $\Delta$. For $\Delta\gtrsim3.744$, the periodic solutions undergo a cascade of period-doubling bifurcations giving rise to chaos with one positive Lyapunov exponent at  $\Delta\approx3.75$  [Fig. \ref{fgr:map}c]. The period-doubling bifurcations do not affect the spectrum of Lyapunov exponents of the stable solution, since the latter remains periodic. However, each period-doubling causes additional Lyapunov exponent to approach zero [Fig. \ref{fgr:map}c], which manifests itself via doubling of the number of dots in Fig. \ref{fgr:map}d. Thus for $N=5$, the bifurcation mechanism leading to the onset of chaos stays the same as for the case of $N$=2. Further investigation shows that this mechanism is also present at larger $N$. 

Increasing $\Delta$ makes the chaotic oscillations more complicated and leads to the gradual development of hyperchaos, which emerges at $\Delta\approx4.4$ [see Fig. \ref{fgr:map}b]. This transition is linked to further instability of already unstable periodic orbits, which form the skeleton of the chaotic attractor. Accumulation of the corresponding bifurcations causes the second Lyapunov exponent to become positive. 

We find that period-doubling is the most common, but not the only, route to chaos exhibited by the system. Figure \ref{fgr:map}a reveals some direct transitions from the red region where the dynamics is quasiperiodic to the green region where the dynamics are chaotic. This behaviour can indicate the mechanism of torus destruction for the transition to chaos . However, we did not specifically investigate the effect of this particular mechanism on the development of hyperchaos. 

\begin{figure*}[t]
\centerline{\includegraphics*[scale=.25]{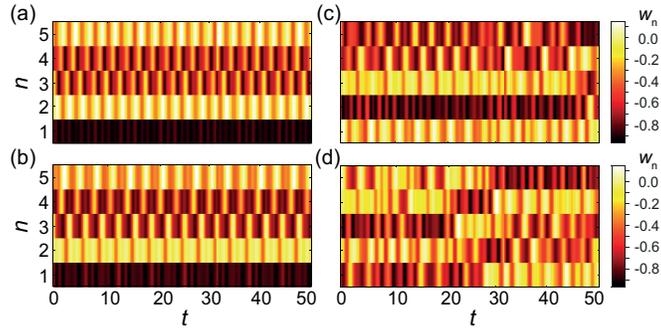}}
	\caption{\textbf{Time evolution of $w_n$ for 5 qubits.} \textbf{a} Periodic [$\Delta=3.74$], \textbf{b} quasiperiodic [$\Delta=3.73$], \textbf{c} chaotic [$\Delta=4.05$] and \textbf{d} hyperchaotic  [$\Delta=4.95$] oscillations; $n$ labels an qubit in the chain.}
	\label{fgr:reall}
\end{figure*}

The dynamics of $N=5$ coupled qubits in different dynamical regimes is exemplified in Fig. \ref{fgr:reall}.
Typical periodic solution is presented in Fig. \ref{fgr:reall}a. Here, $n =1,\ldots,5$ denotes the qubit number in the chain, $t$ is time, and the color scale indicates the value of $w_n$. All qubits oscillate with the same frequency, but with different phases. However, the phase shift is constant for each pair of qubits, meaning that they are synchronized. The latter is also indicated by the clearly periodic pattern in Fig. \ref{fgr:reall}a. The quasiperiodic regime  when $\Delta=3.73$ is shown in Fig. \ref{fgr:reall}b. Now the phase shifts between oscillations of different qubits are no longer constant, and the periodic pattern of the spatio-temporal dynamics is lost, reflecting the loss of synchronization.

Chaotic oscillations for $\Delta=4.05$ are illustrated in  Fig. \ref{fgr:reall}c. Here, the qubit chain demonstrates erratic behavior, which is in marked contrast to the ordered dynamics presented in Figs. \ref{fgr:reall}a and b.
Figure \ref{fgr:reall}d shows typical hyperchaotic oscillations calculated for $\Delta\approx4.95$. In this regime, the oscillations in the chain become even more complicated than the chaotic behavior in Fig. \ref{fgr:reall}c, and now demonstrate no specific time-scales in $w_n(t)$. The hyperchaotic behavior persists up to $\Delta\approx8.0$, after which it rapidly switches to chaotic and then periodic solutions. For $\Delta>8.53$, all oscillations disappear, and all long-term solutions in the system correspond to stable equilibrium.

In order to further examine the presence of the multistability in the dynamics of the chain, we return to the evolution of the Lyapunov exponents as $\Delta$ changes from 9 down to 1. Values of $\Lambda_{1,2,3,4}$ for this case are shown red in  Fig. \ref{fgr:map}b. The plot shows that the stable homogeneous steady state ($\Lambda_1=$ $\Lambda_2=$ $\Lambda_3=$ $\Lambda_4$$<0$) exists down to $\Delta \approx5.491$, where it suddenly changes to hyperchaotic oscillations. Thus, within the interval of $\Delta$ between 5.491 and 9.0 the homogeneous fixed point co-exists with different inhomogeneous regimes including other types of equilibria, periodic and quasiperiodic oscillations, chaos and hyperchaos [blue graphs in Fig. \ref{fgr:map}b]. Multistability therefore appears to be a generic phenomenon, existing in chains of different size.

\begin{figure*}[t]
	%\vskip 4cm
	\centerline{\includegraphics*[scale=.9]{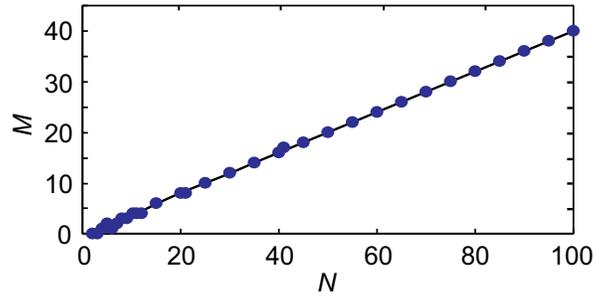}}
	\caption{\textbf{Number of positive Lyapunov exponents $M$ {\it vs.} number of  qubits $N$ in the ring chain.} $\Delta=5$, $\Omega=2.5$, $c=5$.}
	\label{fgr:NLyap}
\end{figure*}

Analysis of chains with $N>5$ reveals that hyperchaos is not only preserved in the system, but becomes more complicated as more Lyapunov exponents become positive. The results of our analysis are summarized in Fig. \ref{fgr:NLyap}, where the number of positive Lyapunov exponents, $M$, is shown as a function of the number of qubits in the chain. The graph shows an almost linear growth, at a rate suggesting that adding two or three qubits leads to the appearance of an additional positive Lyapunov exponent. This phenomenon originates from a weak correlation between the oscillations in distant qubits. As a result, the addition of a subsystem, comprising two or more qubits is able to demonstrate chaotic dynamics and adds one more positive value to the spectrum of the Lyapunov exponents. Since, for a given coupling, the smallest chaotic subsystem comprises two qubits, their inclusion produces one more  positive Lyapunov exponent. The number of Lyapunov exponents grows roughly proportionally to the number $N$ of qubits rather than with the dimensionality $2^N$ of the Hilbert space of the system. This indicates the possibility of a reduced description of its dynamics. In such a case the transitions between qualitatively different, distinct dynamical regimes are controlled by a relatively small number of independent dimensionless parameters.
 
%, see  the Supplementary Materials \cite{Bal19supp} for  details. 
For large $N$, multiple positive Lyapunov exponents make the oscillations very complicated, demonstrating broadband continuous spectra, which are similar to random fluctuations in solids \cite{Field1995,Kogan_book2008}. In addition, we found out that these hyperchaotic phenomena are preserved for non-identical qubits and open chains, as well as in chains with more complex qubit-coupling topology than those discussed above. The spatial correlations, the spectra of the chain oscillations calculated for different numbers of qubits, and Lyapunov analysis of chains with non-identical qubits and more complex topology  are discussed in Sections C and D in Supplementary Materials.
% \cite{Bal19supp}. 

\begin{figure*}[t]
	%\vskip 4cm
	\centerline{\includegraphics*[scale=0.43]{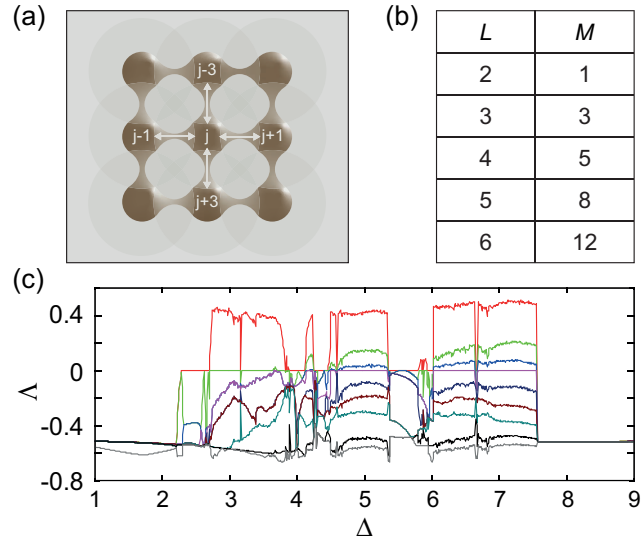}}
	\caption{\textbf{2D $L\times L$ lattices of $L^2$ interacting qubits.} 
		\textbf{a} Schematic representation of a square lattice of nine interacting quantum units (e.g., Rydberg atoms or qubits), labelled by integer indexes and coupled via nearest neighbor interaction. 
		\textbf{b} Number of positive Lyapunov exponents $M$ {\it vs.} lattice size $L$ in the $L\times L$ lattice calculated for $\Delta=5$, $\Omega=2.0$, $c=5$.
		\textbf{c} Variation of the eight largest Lyapunov exponents with $\Delta$ for a square lattice of 9 qubits, $\Omega=2.0$.}
	\label{fgr:2Dlattice}
\end{figure*}

To study the significance of the system's dimensionality, we analyze 2D $L\times L$ lattices of $L^2$ interacting qubits (Fig.~\ref{fgr:2Dlattice}a). As for 1D rings, these square lattices exhibit chaotic (for $2\times2$ lattices) and hyperchaotic behavior (for $3\times3$ and larger lattices). Figure~\ref{fgr:2Dlattice}c illustrates the Lyapunov exponents spectrum for a $3\times 3$ square lattice. We observe two areas ($2.7<\Delta<3.8$ and $\Delta\approx5.8$) of chaos, one small area of hyperchaos with two positive Lyapunov exponents ($\Delta\approx4.2$), and two areas of hyperchaos with three positive exponents ($4.4<\Delta<5.3$ and $6.05<\Delta<7.6$). Therefore for a lattice of 9 qubits we observe a complicated behavior with three positive Lyapunov exponents, which is similar behavior to that we observe for a ring of 9 qubits. We investigate   lattices with several values of $L$ of nodes and and find almost linear growth of the number of positive Lyapunov exponents with increasing the number of qubits in the lattice (see Fig.~\ref{fgr:2Dlattice}b). We can conclude that emergence of hyperchaos does not depend on the dimensionality of the system.

\subsection{Chaos control via a coherent driving field}

\begin{figure*}[t]
	%\vskip 4cm
	\centerline{\includegraphics*[scale=0.9]{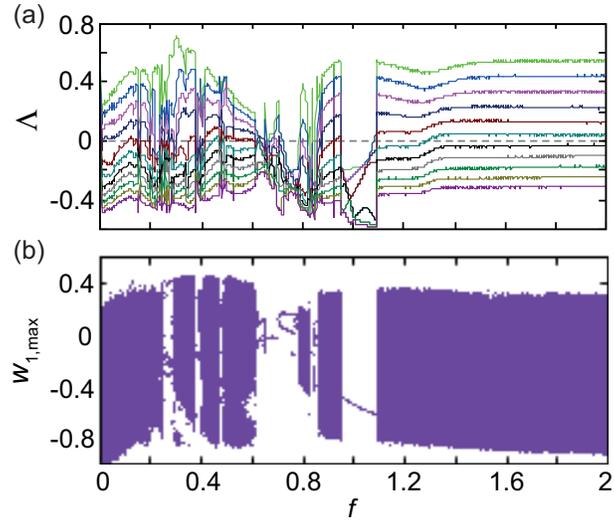}}
	\caption{\textbf{Chaos control.} \textbf{a}  Conditional Lyapunov exponents spectrum and \textbf{b} bifurcation diagram for a ring of 15   qubits by means of an external parametric effect for $\Omega_m=2.5$, $c=5$, $\Delta=5.0$, $M=0.684$.}
	\label{fgr:LyapExpExtChangef}
\end{figure*}

This dynamical complexity has a deterministic origin, which could be controlled (suppressed or enhanced) by an applied external field. Previously, it was demonstrated that a periodic perturbation can suppress hyperchaos with two positive Lyapunov exponents \cite{Xiao_hui_1999, Sun2012}. Here, we apply a periodic modulation to the laser field amplitude which modulates the Rabi frequency: 
\begin{equation}\label{eq:ExtEff}
\Omega=\Omega_{m}[1+M\sin(2\pi ft)].
\end{equation}
where $\Omega_{m}$ is the amplitude of the Rabi frequency, whilst $M$ and $f$ are the modulation index and the frequency of modulation, respectively. 
We consider the case when $N=15$, a large but practically feasible number of qubits such that the system shows hyperchaos, characterized by four positive Lyapunov exponents and a broadband spectrum. %\cite{Bal19supp}.  
Figure \ref{fgr:LyapExpExtChangef} presents the eleven largest conditional Lyapunov exponents (a) and the corresponding bifurcation digram (b) calculated for $\Omega_m=2.5$, $c=5$, $\Delta=5.0$, $M=0.684$ and $f$ changing from 0 to 1.8.

Due to the periodic modulation, chaos is suppressed in certain parameter regiones. For $f$ between 0.7 and 0.8 we find complex periodic oscillations characterized by the largest Lyapunov exponent less then 0 [see Fig. \ref{fgr:LyapExpExtChangef}a] and multiple branches in the bifurcation diagram [Fig. \ref{fgr:LyapExpExtChangef}b]. Within the interval $f\in (0.9, 1.1)$ we find period-one oscillations (one maxima per period) and Lyapunov exponents $\ll0$, corresponding to highly stable regular oscillations in this regime. In addition, the same controlling signal is able to significantly increase the complexity of hyperchaotic oscillations and a corresponding increase in the value of the largest Lyapunov exponent. For example, when $f\approx0.3$ the largest Lyapunov exponent becomes almost three time larger than in the case without the application of the signal ($f=0$).  Otherwise the number of positive Lyapunov exponents can be increased by 1 ($f=1.2$) or 2 ($f=1.4$). Our results demonstrate that periodic modulation of laser field amplitude can be used as an efficient method to control very complex hyperchaos in  qubit arrays.

\section{Discussion}
We have shown that the interplay between dissipation and energy pumping in quantum systems comprising chains of qubits can produce highly nontrivial emergent phenomena associated with the onset of complex chaotic and  hyperchaotic oscillations even in the absence of multi-particle entanglement. The complexity of the hyperchaos increases with the number of elements in the chain. 
The number of positive Lyapunov exponents grows linearly with the number of qubits, that is, as only log of the dimensionality of the Hilbert space. This indicates the possibility of a reduced description of the quantum system by a  small  (in comparison to the dimensionality of the Hilbert space) number of dimensionless parameters. This observation is consistent with the   fact that the manifold of all quantum  states that can be generated by arbitrary time-dependent local Hamiltonians in a polynomial time occupies an exponentially small volume in the Hilbert space of the system\cite{Poulin2011}.

Our results demonstrate a mechanism for randomizing the evolution of coupled qubits, which arises due to dynamical phenomena far from equilibrium and is thus unrelated to thermalization processes despite the superficial similarity. The model we have used is generic and can be applied and implemented directly in, e.g., chains of   qubits or electromagnetically driven superconducting qubits. Our results will be important for the development of large quantum systems, where multi-point entanglement is neither required nor supported \cite{Zagoskin2009a, Macha2014, Fitzpatrick2017}. In particular, they suggest a controllable way of switching between different dynamical regimes via regular to chaos transitions either by tuning the parameters of the system or by applying a controlling signal via modulation of laser field amplitude, as another approach towards controlling quantum state dynamics \cite{Zhang2017}. Our results are of interest for the development of quantum random number generators \cite{Herrero2017}  and quantum chaotic cryptography \cite{Monifi2016,deOliveira2018}. A natural extension of this research will consider the effect of interqubit entanglement on the complexity revealed here, and whether it can be utilised for controlling this far-from-equilibrium behavior. 

Another interesting task will be to study how quantum signatures of classical chaos such as the growth rate of out-of-time-ordered correlators \cite{Carlos2019} behave in the presence of hyperchaos. Preliminary results on the integration of the original quantum model (\ref{eq:rho}), briefly discussed in Section E in Supplementary Materials, indicate that the complexity of chain dynamics depends highly non-trivially on the parameters of the Hamiltonian (\ref{eq:Ham}). However, a detailed analysis of the dynamical regimes and their characteristics in the presence of entanglement, and especially of the question about the universality of these regimes and parameters that control them, requires extensive further research, which we hope this paper will stimulate.

\section{Methods}
\subsection{Lyapunov exponents calculation}
The Lyapunov exponents were calculated for the stable limit sets, which correspond to the solutions of the model equations as time $t\rightarrow\infty$. 
Since we apply the periodic modulation (\ref{eq:ExtEff}) to the system (\ref{eq:3}), the Lyapunov exponents we calculate in this case are called conditional and missing one zero exponent, unlike the common Lyapunov exponents spectrum \cite{pyragas1997conditional}.
To determine them, we implement $3N$ perturbations vectors, where $N$ is the number of units in the system, and use periodic Gram-Schmidt orthonormalization of the Lyapunov vectors to avoid a misalignment of all the vectors along the direction of maximal expansion \cite{Oseled68,Wolf_1985}.

\subsection{Bifurcation diagrams}
In order to construct a bifurcation diagram we plot the points corresponding to the local maxima $w_{1,max}$ in the time-evolution of $w_1(t)$, calculated for given $\Delta$ and $\Omega$.

\section{Data availability statement}

The data that support the findings of this study are available from the corresponding authors upon reasonable request.

\section{Code availability statement}

All code used in the paper are available from the corresponding authors upon reasonable request.

\section{Acknowledgements}

A.M.Z. and A.G.B. thank Prof. S. Watabe for illuminating discussions.
This work was supported in part by EPSRC 
%grant ``Testing quantumness: from artificial quantum arrays to lattice spin models and spin liquids" 
(grant EP/M006581/1), the grant from the Ministry of Science and Higher Education of the Russian Federation in the framework of Increase Competitiveness Program of NUST MISiS, Grant No.~K2-2020-001 and Russian Foundation for Basic Research (RFBR) 18-32-20135. W.L. acknowledges support from the UKIERI-UGC Thematic Partnership No. IND/CONT/G/16-17/73, EPSRC Grant No. EP/R04340X/1, and support from the University of Nottingham.

\section{Author contributions}
A.M.Z., A.G.B. and A.E.H. conceived, designed and supervised the project. The theoretical	calculations were performed by A.M.Z. (analytical) and A.V.A., V.V.M., A.E.H. and A.G.B. (numerical). A.G.B., T.M.F., M.T.G., W.L. and A.M.Z. wrote the manuscript. All authors discussed the results and contributed to the manuscript.
%}

\section{Competing interests}
The authors declare no competing interests.

%\bibliography{REFERENCES2020}
%\bibliography{Review-K}

%\bibliographystyle{naturemag}

\end{document}